\documentclass[10pt, twocolumn]{revtex4-2}
\usepackage{graphicx}
\usepackage{subcaption}
\usepackage{amsmath}
\usepackage{enumitem}
\usepackage{braket}
\usepackage{float}
\usepackage{bbold}
\usepackage{hyperref}

\newlist{subenumerate}{enumerate}{2}
\setlist[subenumerate,1]{label=\arabic{subenumeratei}.\arabic*}
\setlist[subenumerate,2]{label=\arabic{subenumeratei}.\arabic{subenumerateii}}

\setlist[enumerate]{label=\arabic*.}

\usepackage[T1]{fontenc}
\usepackage{caption}

\begin{document}
\onecolumngrid
\begingroup  
\centering
\Large \textbf{Optimizing Resource Costs: A Practical Guide to Achieving Target Security in Verifiable Blind Quantum Computing}\\[1.5em]
\large J van Dam$^{1,2,3,*}$, M van Hooft$^1$ and SDC Wehner$^{1,2,3}$\\

\endgroup
\vspace{10pt}
\small \noindent \par
$^1$ QuTech, Delft University of Technology, Lorentzweg 1, 2628 CJ, Delft, The Netherlands\\
$^2$ Kavli Institute of Nanoscience, Delft University of Technology, Lorentzweg 1, 2628 CJ, Delft, The Netherlands\\
$^3$ Quantum Computer Science, EEMCS, Delft University of Technology, Lorentzweg 1, 2628 CJ, Delft, The Netherlands\\
$^*$ Corresponding author: j.vandam-3@tudelft.nl\\

\begin{abstract}
    Verifiable blind quantum computing (VBQC) enables a resource-limited client to securely delegate computations to an untrusted quantum server while maintaining privacy and detecting deviations from the prescribed computation. The noise-robust VBQC protocol of Leichtle et al. achieves this through a round-based structure: the client delegates multiple computation rounds and test rounds, using the test outcomes to detect cheating while tolerating honest hardware noise. The protocol's security proof involves numerous interdependent parameters, making it non-trivial to find a valid parameter set for a given hardware noise level and security target. We formalize this as a constrained optimization problem and develop a practical framework to solve it. The framework yields the protocol parameters that minimize the number of rounds for any given setup. We derive a heuristic formula for the minimal number of rounds to help understand the scaling with noise and security targets and to provide rapid resource estimation. Since the number of rounds depends on noise while the time per round depends on hardware rate, the framework also enables optimization of rate-fidelity trade-offs to minimize end-to-end runtime. We demonstrate both applications through a case study of a trapped-ion server with a measurement-only client, showing how the client's polarization control hardware specifications translate into protocol parameters and runtime estimates, providing concrete guidance for near-term implementations.
\end{abstract}
\maketitle

\section{Introduction}
Cloud-based quantum computing could greatly increase the accessibility of quantum computing, but users have to worry about two things: how do you run sensitive computations on someone else's quantum computer, without exposing your data, and how do you trust the result? These questions have motivated the development of blind quantum computing (BQC) protocols \cite{childs2001secure, broadbent2009universal}, with verifiable blind quantum computing (VBQC) \cite{gheorghiu2019verification} allowing a resource-limited client to execute a quantum algorithm on a powerful, untrusted, remote quantum server. A VBQC protocol ensures that the client's input, computation and output remain private (blind), while also being able to detect any deviation from the prescribed computation (verifiable).\\
Both BQC and VBQC have moved beyond theory: BQC has been demonstrated on photonic platforms \cite{barz2012demonstration, greganti2016demonstration, huang2017experimental}, verification of a delegated computation was demonstrated shortly after \cite{barz2013experimental}, and, most recently, a full VBQC protocol was executed on a trapped-ion server networked to a client over a fibre link \cite{drmota2024verifiable}. These are proof-of-principle demonstrations: they establish that the protocols run on real hardware, typically for small computations and without optimizing the resources needed to reach a chosen security level. Determining those resources is precisely the gap we address, and is a prerequisite for turning such demonstrations into deployable implementations.\\
Verifiability of the computation can be achieved in different ways, such as either trap based \cite{vbqc}, stabiliser based \cite{hayashi2015verifiable} or classical with computational assumptions \cite{mahadev2018classical}. A crucial aspect of the security proof is noise-robustness: In the noisy intermediate-scale quantum (NISQ) \cite{Preskill2018quantumcomputingin} regime, noise-robustness allows honest noise to be tolerated without compromising statistical security, while still accepting (not aborting) with high probability. In the fault-tolerant quantum computing (FTQC) regime, noise-robustness requires demonstrating that the protocol can be executed securely at the encoded level using fault-tolerant gadgets that are both correct and secure \cite{kapourniotis2025plugging}. To this end, the noise-robust VBQC (rVBQC) protocol by Leichtle et al. was introduced \cite{leichtle2021verifying} for bounded-error quantum polynomial time (BQP) computations.\\
The rVBQC protocol achieves its noise robustness through a round-based structure. The client delegates a total of $n=d+t$ rounds to the server: $d$ rounds are used to execute the actual computation, while $t$ rounds are used to test the server. In a test round, the client inserts `trap' qubits, which the client asks the server to measure in their eigenbases, so that the correct outcome (in a noiseless case) is known to the client. From the server's perspective, the computation rounds and test rounds are indistinguishable, hence it cannot cheat selectively. At the end of the protocol, the client counts how many test rounds produced at least one unexpected trap outcome; if this exceeds a threshold $\omega$, the client aborts, otherwise it accepts. When the client accepts, the outcome of the computation is taken to be the majority vote over computation rounds. The protocol's noise tolerance is captured by the parameter $p_\text{max}$: an upper bound on the probability that at least one trap measurement fails within any single test round. Leichtle et al. show that, for $p_\text{max}$ below a certain threshold determined by the computation, there exists a set of protocol parameters such that the scheme accepts with high probability on honest-but-noisy devices while maintaining negligible security and correctness errors.\\
However, translating this theoretical security of the protocol into practice presents significant challenges. While the rVBQC protocol provides the theoretical key to secure computation on noisy devices, practical implementation requires resource optimization. Determining the total number of rounds $n$ required to achieve a target security level $\epsilon_\star$ is non-trivial. The value of $n$ depends on the noise in the system (which determines $p_\text{max}$) as well as several free ``tuning parameters'' that control the security proof. The most significant of these parameters are directly physical, they define the protocol's structure, such as the fraction of rounds dedicated to computation versus testing ($\tau=t/n$), and the client's operational strategy, like the chosen tolerance for failed test rounds ($\omega$). The others are more subtle; they control how the security proof bounds random fluctuations, in the number of corrupted rounds, triggered traps, and so on, that determine whether a cheating strategy can succeed undetected.\\
Leichtle et al.~\cite{leichtle2021verifying} note, in their analysis of the number of repetitions, that the optimal ratio of computation to test rounds can be obtained numerically from their security bounds. Separately, Kapourniotis et al.~\cite{kapourniotis2024unifying} optimise the \emph{design} of the test rounds: by establishing a correspondence between verification and error-detection, they characterise which test constructions a client may mix into the computation, and how that choice governs the resource overhead and noise tolerance of the protocol. What neither provides is a concrete operating point. Given a hardware
noise level $p_\text{max}$ and a target security $\epsilon_\star$, we jointly optimise \emph{all} of the protocol's tuning parameters, not only the computation-to-test ratio $\tau$, but also the abort threshold $\omega$ and the abstract parameters that control the security proof, to minimise the total number of rounds $n$, derive a heuristic for $n(p_\text{max}, \epsilon_\star)$, and translate it into an end-to-end
runtime.\\
Formalizing and solving this optimization problem will provide a guide for implementation: it provides the best strategy for the client. This strategy includes how many rounds to execute, how many of those rounds are used for testing vs how many for computation and how many of the tests are allowed to fail before the client aborts. To help get an understanding of the scaling of the number of rounds, and to provide a tool for rapid resource estimation, we deduce a heuristic function to find $n(p_\text{max}, \epsilon_\star)$.\\
Beyond providing an implementation guide, our framework also addresses a more general challenge: quantum network systems often exhibit rate-fidelity trade-offs \cite{coutinho2023entanglement, tanji2024rate, hermans2023entangling, van2025single}. Because our optimization, among other things, yields $n$ as a function of $p_\text{max}$, we quantify how the number of rounds increases with noise. This allows us to find a minimal end-to-end runtime that balances the time-per-round (determined by the rate) with the number of rounds (determined by the `fidelity', or noise level): identifying an optimal point in the rate-noise curve. \\
We apply this framework to a case study of a trapped-ion quantum server and a measurement-based client to give an example of what an optimal client strategy for this setup looks like, to provide concrete end-to-end runtime estimates and to study a realistic trade-off scenario. In this setup, the client controls the state of the servers ions by performing remote state preparation. The client measures a ion-entangled polarization-encoded photon in a basis of its choice, thereby projecting the ion state onto that same basis. In order to measure the photon in an arbitrary basis, the client must switch between polarization bases between each attempt to keep fresh randomness. We analyze a trade-off associated with the polarization control: faster switching often comes with less angular precision and vice versa. We show that hardware with faster switching times can tolerate higher angular errors while still achieving target runtimes, this provides concrete guidance for hardware selection in near-term implementations.\\
Thus, our contributions are:
\begin{itemize}
    \item Formalizing the round-minimization problem in rVBQC as an optimization problem grounded in the security proof by Leichtle et al. \cite{leichtle2021verifying}, and developing a practical framework to solve it.
    \item Deriving a heuristic function for this minimal number of rounds, enabling quick resource estimation.
    \item Showing how rate–fidelity trade-offs can be optimized to minimize end-to-end runtime of the rVBQC protocol in any client-server quantum network system exhibiting such trade-offs.
    \item Illustrating the complete framework through a case study, demonstrating how hardware choices (here, the polarization control hardware of the client) translate into protocol parameters and end-to-end runtime.  
\end{itemize}

This paper is organized into two parts. Part I describes the optimization problem and general results. Part II applies the framework to a case study of a trapped-ion server with a measurement-only client.

\part{The optimization problem}
Before formalizing the optimization problem, we first introduce some preliminaries to understand the rVBQC protocol by Leichtle et al. and the basics of the protocol's security proof. We then formalize the problem, explain our method of solving and provide general solutions.

\section{Preliminaries}
To fully understand the optimization problem, some background knowledge is required. Here, we aim to give the required background in both the protocol and the security proof of \cite{leichtle2021verifying} and to provide context for the optimization problem. For a full understanding we invite the reader to work through the protocol and security proof of \cite{leichtle2021verifying} as well as the cited literature themselves.
\subsection{Blind quantum computing basics, verification basics and the rVBQC protocol}
To understand the optimization problem at the heart of this work, we first review the foundational concepts of the underlying protocol: the Measurement-Based Quantum Computation (MBQC) model, the principles of blindness and verification, and finally, the specific noise-robust protocol that we analyse.
\subsubsection{Measurement-based quantum computing}
The rVBQC protocol relies on the MBQC model \cite{raussendorf2001one}, an alternative to the more common circuit-based model. In MBQC, the computation is driven by single-qubit measurements performed on a highly entangled multi-qubit resource state, known as a cluster or graph state. A graph state is a quantum state that directly represents a mathematical graph $G = (V, E)$, where each vertex $v \in V$ corresponds to a qubit initialized in the $|+\rangle$ state, and each edge $e \in E$ corresponds to a controlled-Z (CZ) gate applied between the respective qubits \cite{hein2004multiparty}. The entire computation is then carried out by performing a sequence of single-qubit measurements on the graph state. Each measurement is performed in a specific basis, defined by an angle on the equatorial plane of the Bloch sphere, and the choice of these angles determines the computation that is executed.\\

\subsubsection{Blind delegation in the MBQC model}
Consider a scenario where a resource-limited client wants to execute a quantum computation. Because the client does not have the resources themselves, they delegate to a remote server. If the client wants their computation to remain private, it can use blind quantum computing (BQC) \cite{broadbent2009universal}. In BQC, the server remains ignorant about the client's input, output and computation, it only learns an upper bound on the size of the computation. The computation can be hidden by hiding the measurement angles that define the MBQC computation from the server. The client can achieve this by adding random rotations to the qubits in the graph, i.e., instead of asking the server to prepare the qubits in the standard $|+\rangle$, the client remotely prepares them in a rotated state $|+_\theta \rangle = (|0\rangle + e^{i\theta} |1\rangle)/\sqrt{2}$ \cite{bennett2001remote}, where $\theta$ is a random angle only known to the client. To preserve the computation while hiding the true measurement angle, the client instructs the server to measure at an adjusted angle that compensates for the random rotation $\theta$. Since the server knows only this adjusted angle, not $\theta$ itself, the true measurement angle remains secret.

\subsubsection{Verification}
In addition to keeping the computation secret, we also want a way to check that the server performs as it is supposed to. This is where verification comes in. Verification is commonly achieved by embedding "traps" within the computation \cite{vbqc}. A trap is a qubit for which the client can deterministically determine the outcome when it is measured in a specific basis, such that the client can detect deviations with what the server reports.\\
A way to create these traps is by surrounding a qubit in the graph state by "dummy qubits". A dummy qubit is a qubit prepared in the computational basis ($|0\rangle$ or $|1\rangle)$. A qubit all of whose neighbours are dummy qubits is effectively decoupled from the rest of the graph (i.e., it is not entangled), and therefore becomes a stand-alone $|+_\theta \rangle$ qubit. The client can then ask the server to measure such a trap qubit in its eigen-basis, hence the client will know what the outcome of the measurement is supposed to be. A key principle of this security model is that, from the server's perspective, the instructions for preparing and measuring a trap are indistinguishable from those for a computation qubit, preventing it from cheating selectively.
\subsubsection{The rVBQC protocol}\label{subsec:rvbqc}
The work by Leichtle et al. \cite{leichtle2021verifying} introduces a noise-robust protocol that modifies the trap structure to enhance practicality and noise tolerance. Instead of embedding individual traps within a single, large computational graph, this protocol separates the $n$ total rounds into two sets. The full protocol consists of $d$ computation rounds, which are used to run the actual BQC algorithm, and $t$ test rounds, which are composed entirely of trap and dummy qubits. Since each trap needs to be surrounded by dummies, the maximal traps that can be embedded in a given graph is determined by the graph's chromatic number. Coloring is a method of assigning labels (colors) to vertices in a graph so that no two adjacent vertices share the same color. If the graph is $k$-colorable, i.e., it has chromatic number $k$, we can color the graph using $k$ colors, which means $1/k$ of the qubits in the graphs can become traps. This affects the probability that a deviation in the server's implementation is caught by the client.\\
The separation of trap and computation rounds lowers the state preparation requirements for the server in any given round. More importantly, it provides a powerful mechanism for noise tolerance. A real-world quantum device will have a non-zero physical error probability, leading to non-zero probability of a test round failing ($p_\text{max}$) even with an honest server. To accommodate this, the protocol does not demand that all test rounds pass. Instead, the client sets a failure tolerance, $\omega$, which is the maximum fraction of test rounds that are allowed to fail before the entire protocol is aborted. This makes the protocol viable as long as the hardware's error rate is below this threshold ($p_\text{max} < \omega$). The final result of the algorithm is then determined by a classical majority vote over the outcomes of the $d$ computation rounds, which provides robustness against random errors that occur during the computation.\\
Now that we have an idea of what the protocol looks like, we can look into why it is secure. In the next section, we give a brief overview of the security proof for this protocol, which we then use to formalize our optimization problem.

\subsection{Security proof outline}\label{subsec:secproof}
With the protocol structure established, we now examine how its security is formally proven in Ref.~\cite{leichtle2021verifying}. The security of the rVBQC protocol is formally established within the composable framework of Abstract Cryptography \cite{maurer2011}, which guarantees that the protocol remains secure even when used as a component in larger systems. In this framework, security is defined by the indistinguishability between the real protocol (with noise and potentially dishonest parties) and the ideal functionality that always produces the correct result. The overall security level $\epsilon$ quantifies the maximum probability that anyone can tell the real from the ideal functionality. The protocol achieves $\epsilon$-composable security with $\epsilon=\text{max}\{\epsilon_\text{sec},\epsilon_\text{cor}\}$, where $\epsilon_\text{sec}$ captures security against a malicious server and $\epsilon_\text{cor}$ captures correctness on honest-but-noisy devices.\\
Below, we will give some intuition on how the protocol achieves this $\epsilon$-composable security, after which we present the formal criteria which will later be used to formalize our optimization problem.\\
The first important insight is that if the protocol is blind, that blindness forces any attack to affect computation and test rounds equally, because the server \footnote{In this context, the `server' is anything but the client itself, so any evesdropper on the channel, for example, is absorbed, from a security standpoint, into the server.} cannot distinguish them. Therefore, if we measure a certain deviation strength in the test rounds, this same deviation strength applies to the computation rounds, up to statistical fluctuations.\\
Because of the majority vote over the computation rounds and the bounded-error nature of BQP, the server must successfully corrupt at least a fraction $Z=(2p-1)/(2p-2)$ of all rounds to flip the total result, where $0\leq p < 1/2$, with $p$ the inherent error probability of the BQP algorithm. A cheating server then has to decide between:
\begin{enumerate}
    \item Attacking conservatively: Attacking fewer than a $Z$ fraction of rounds on average, avoiding detection with high probability but also flipping the majority vote with low probability;
    \item Attacking aggressively: Attacking at least a $Z$ fraction of rounds on average, to flip the majority vote with higher probability but also have a low probability of doing this undetected.
\end{enumerate}
The security proof bounds both scenarios. Each bound can be tuned by independently adjusting how much statistical slack we allow around expected values for cheating strategies.\\
Following the reduction of Dunjko et al. \cite{dunjko2014composable}, the composable security of any verifiable delegated quantum computation protocol can be established through four stand-alone criteria:
\begin{enumerate}
    \item $\epsilon_\text{cor}$-local-correctness: The probability that the protocol fails to deliver the correct answer to the client is bounded by $\epsilon_\text{cor}$.
    \item $\epsilon_\text{bl}$-local-blindness: The server's state at the end of the protocol is indistinguishable from what it could have generated on its own, ensuring the client's input, computation, and output remain private.
    \item $\epsilon_\text{ver}$-local-verifiability: The probability that a malicious server causes the client to accept an incorrect result is bounded by $\epsilon_\text{ver}$.
    \item $\epsilon_\text{ind}$-independent-verification: The server can determine on its own, using the transcript of the protocol and its internal registers, whether or not the client will abort.
\end{enumerate}
When these conditions are satisfied, the protocol is $\epsilon$-composably secure with $\epsilon=\text{max}\{\epsilon_\text{sec},\epsilon_\text{cor}\}$. Here, the security parameter is given by 
\begin{equation}\label{eq:sec}
    \epsilon_\text{sec}=4\sqrt{2\epsilon_\text{ver}}+2\epsilon_\text{bl}+2\epsilon_\text{ind}.
\end{equation}
For our purposes, the security analysis simplifies significantly as two of these criteria are perfectly satisfied by the protocol's design. The blindness is perfect ($\epsilon_\text{bl}=0$). This is because the client uses a one-time pad for the measurement angles, and the server-side operations for a computation round are statistically indistinguishable from those for a test round. Even in the case of an abort, the information revealed to the server (e.g., that a specific round contained a trap) is uncorrelated with the client's private input or the logic of the computation. Similarly, independent verification is also perfect ($\epsilon_\text{ind} = 0$). The protocol concludes with the client explicitly sending an "Ok" or "Abort" message, so the server knows the final outcome of the verification process without ambiguity. With these simplifications:
\begin{equation}\label{eq:sec_ver}
    \epsilon_\text{sec}=4\sqrt{2\epsilon_\text{ver}} \implies \epsilon_\text{ver}=\epsilon_\text{sec}^2/32.
\end{equation}
The security analysis therefore reduces to bounding two quantities: the correctness error $\epsilon_\text{cor}$ (which includes both rejection due to noise and computational errors) and the verifiability error $\epsilon_\text{ver}$ (the probability a malicious server escapes detection while corrupting the result).

\section{Optimization problem and solving methods}
We now state the optimization problem formally. Let $\mathbf{x}=(\phi, \epsilon_1, \epsilon_2, \epsilon_3, \tau)$ denote the vector of tuning parameters. The problem is: minimize $n(\mathbf{x}) = \max (n_\text{cor}(\mathbf{x}), n_\text{sec}(\mathbf{x}))$, subject to
\begin{enumerate}
    \item $0<\phi<Z$
    \item $0<\epsilon_1<Z-\phi$
    \item $0<\epsilon_2<1/k$
    \item $0<\epsilon_3<\phi$
    \item $0<\tau<1$
    \item $\omega(\mathbf{x})>p_\text{max}$
\end{enumerate}
where $n_\text{cor}$ and $n_\text{sec}$ are given by Equations \ref{eq:ncor} and \ref{eq:nsec}, and $\omega$ is given by Equation \ref{eq:omega}.\\
To solve this optimization problem, we derive two constraints, formulate an objective function and provide a method for solving this objective function. This involves a lot of variables that are introduced in close succession, to help the reader keep track of the meanings, a glossary is provided in Appendix \ref{app:glossary}.

\subsection{Constraints}
From $\epsilon=\text{max}\{\epsilon_\text{sec}, \epsilon_\text{cor}\}\leq\epsilon_\star$ we require both $\epsilon_\text{sec}\leq\epsilon_\star$ and $\epsilon_\text{cor}\leq\epsilon_\star$. \\
\textbf{Correctness.} The correctness error has two components: $\epsilon_\text{cor}=\epsilon_\text{rej}+\epsilon_\text{ver}$, where $\epsilon_\text{rej}$ is the probability of aborting due to honest noise. It is shown in Reference \cite{leichtle2021verifying} that
\begin{equation}
    \epsilon_\text{rej}=\exp\left[-2(\omega - p_\text{max})^2 \tau n \right],
\end{equation}
with $\omega$ the fraction of allowed failed test rounds, $p_\text{max}$ the probability of failing a test round due to noise, $\tau$ the fraction of rounds used for testing, and $n$ the total number of rounds.\\
Combined with Equation \ref*{eq:sec_ver} our correctness requirement becomes
\begin{equation}\label{eq:testreq1}
    \exp\{-2(\omega - p_\text{max})^2\tau n\} \leq \epsilon_\star - \epsilon_\star^2/32.
\end{equation}

\textbf{Security.} Recall from Section \ref{subsec:secproof} that a cheating server can either attack conservatively (attack less than $Z$ fraction of rounds) or aggressively (attack at least $Z$ fraction of rounds). Leichtle et al. bound the failure probability for each strategy using tail bounds on distributions, yielding:
\begin{equation}\label{eq:prfail}
    \begin{split}
    \text{Pr[fail]} \leq \text{max}\{ &\exp(-An) + \exp(-Bn), \\
    &\exp(-Cn) + \exp(-Dn)\},
    \end{split}
\end{equation}
The first term bounds the conservative strategy: $A$ controls the probability that, despite attacking few rounds, enough computation rounds are corrupted to flip the majority vote, while $B$ bounds the probability that the attack strength is unexpectedly concentrated on computation rounds rather than test rounds. The second term bounds the aggressive strategy: $C$ controls the probability that, despite attacking many rounds, few test rounds are hit, while $D$ bounds the probability that among the affected test rounds, few traps are triggered.\\
The exponents (derived in \cite{leichtle2021verifying}) depend on the protocol structure and the tuning parameters that determine the statistical slack:
\begin{equation}
    \begin{split}
        A &= 2(1-Z + \phi - \epsilon_3)\delta\epsilon_4^2,\\
        B &=\frac{2\delta^2\epsilon_3^2}{Z - \phi},\\
        C &= \frac{2\tau^2\epsilon_1^2}{Z - \phi} ,\\
        D&=2(Z - \phi - \epsilon_1)\tau\epsilon_2^2.
    \end{split}
\end{equation}
Here, $\delta=d/n$ is the fraction of computation rounds and $Z=(2p-1)/(2p-2)$. The parameters $\phi, \epsilon_1, \epsilon_2$, and $\epsilon_3$ are the tuning parameters that set the confidence buffers in the tail bounds. The allowed values for these tuning parameters are given in Table \ref{tab:tuning_params}. An additional threshold is defined in \cite{leichtle2021verifying} as
\begin{equation}\label{eq:eps4}
    \epsilon_4=\frac{1/2 - Z + \phi - \epsilon_3}{1 - Z + \phi - \epsilon_3} - p.
\end{equation}
It is guaranteed in combination with other constraints that $\epsilon_4>0$.\\
With these tuning parameters we can also set the threshold for the number of test rounds we allow to fail before aborting, bounded away from its maximum, $Z/k$ (determined by how likely we are to catch deviations):
\begin{equation}\label{eq:omega}
        \omega = (1/k-\epsilon_2)(Z - \phi - \epsilon_1).
\end{equation}
Leichtle et al. establish that $\text{Pr[fail]}\leq \epsilon_\text{ver}$. Combined with Equation \ref{eq:sec_ver}, and our target $\epsilon \leq \epsilon_\star \implies \epsilon_\text{sec}\leq \epsilon_\star$ we then find that we require
\begin{equation}\label{eq:testreq2}
    \text{Pr[fail]}\leq \epsilon_\star^2/32.
\end{equation}

\begin{table}[]
    \centering
    \begin{tabular}{|p{0.5cm}|p{5cm}|p{2.2cm}|}
        \hline 
        & \textbf{Meaning} & \textbf{Allowed interval} \\
        \hline
        $\phi$ & Used to define a threshold for the number of `affected' computation rounds & $0<\phi<Z$\\
        \hline
        $\epsilon_1$ & Used to bound the probability of the number of affected test rounds being significantly lower than expected & $0<\epsilon_1<Z-\phi$\\
        \hline
        $\epsilon_2$ & Buffer used in setting the round failure threshold & $0<\epsilon_2<1/k$\\
        \hline
        $\epsilon_3$ & Used together with $\epsilon_4$ to relate the probability of the number of affected computation rounds exceeding a threshold to the parameters $p$ and $n$ & $0<\epsilon_3<\phi$\\
        \hline
    \end{tabular}
    \caption{Tuning parameters: thresholds that can be varied in the optimization problem}
    \label{tab:tuning_params}
\end{table}

\subsection{Objective function}
We find the objective function by solving the security and correctness requirements: Equations \ref{eq:testreq1} and \ref{eq:testreq2}. Because the functions in Equations \ref{eq:testreq1} and \ref{eq:testreq2} are monotonically decreasing with $n$, we find the minimal $n$ by solving the equality. The objective function should thus return the $n$ that solves the equality for both requirements for a given set of input tuning parameters $\phi, \epsilon_1,\epsilon_2,\epsilon_3, \tau$ and the input parameters set by the problem $k$ (the chormatic number of the graph the computation is carried out on), $p$ (the inherent error probability of the computation), $\epsilon_\star$ (the target level of security) and $p_\text{max}$ (the probability of a test failing in the honest-but-noisy case).\\
Finding the $n$ that satisfies the equality for correctness can be found directly as 
\begin{equation}\label{eq:ncor}
    n_\text{cor}= \frac{-\log(\epsilon_{cor, target} - \epsilon_{sec, target}^2/32)}{2(\omega - p_{max})^2\tau}.
\end{equation}
For the security condition, we have to do a bit more work. First, we can divide the inequality into two. Because we require the maximum of two terms to be below a certain threshold, we need both of those two terms to be below the threshold individually as well. This gives us two inequalities that are again solved minimally by the equality, leading to two conditions:
\begin{enumerate}
    \item $\exp(-An) + \exp(-Bn) = \epsilon_{sec, target}^2/32$,
    \item $\exp(-Cn) + \exp(-Dn) = \epsilon_{sec, target}^2/32$.
\end{enumerate}
To solve these equations for $n$, we use the LogSumExp approximation, which states that for $x,y>0$, we have max($x,y)\leq\log (e^x + e^y)\leq \text{max}(x,y) + \log(2)$. Applying this to our first condition, we find
\begin{equation}
    \frac{\log(32/\epsilon_{\text{sec}}^2)}{\min(A,B)} \leq n \leq \frac{\log(64/\epsilon_{\text{sec}}^2)}{\min(A,B)},
\end{equation}
and analogously for the second condition with $\min(C,D)$. Since both conditions must be satisfied simultaneously, we take the more restrictive bound by replacing $\min(A,B)$ and $\min(C,D)$ by $\min(A,B,C,D)$, giving
\begin{equation}
    \frac{\log(32/\epsilon_\star^2)}{\text{min}(A,B,C,D)} \leq n_\text{sec} \leq \frac{\log(64/\epsilon_\star^2)}{\text{min}(A,B,C,D)},
\end{equation}
The upper and lower bounds differ only by a factor of 2 inside the logarithm. For typical security parameter targets $\epsilon_\star ~ 10^{-3}$ to $10^{-7}$, the correction is only a few percent difference. We therefore use the sufficient (upper) bound as our approximation:
\begin{equation}\label{eq:nsec}
    n_\text{sec} \approx \frac{\log(64/\epsilon_\star^2)}{\text{min}(A,B,C,D)}.
\end{equation}
To satisfy both security and correctness, we need to have the amount of rounds that is at least the maximum of the security and correctness minima:
\begin{equation}\label{eq:n}
    n = \text{max}(n_\text{cor}, n_\text{sec}).
\end{equation}
With the objective function in place, we now turn to methods for solving the optimization problem.

\subsection{Solving method}
Since our objective function includes minimum and maximum operations, the problem landscape is not smooth. Because of this, numerical optimization methods based on gradients will struggle. In addition, the optimization problem is constrained as various parameters are bound by other parameters, e.g., $\epsilon_1$ and $\epsilon_2$ have a maximum that depend on $\phi$. We have found that differential evolution \cite{storn1997differential} is a suitable solver for this problem, because it is a derivative-free global optimizer that handles constraints well and is easy to use in its scipy implementation \cite{virtanen2020scipy}. The code that handles this can be found in \cite{code}.

\section{General results and discussion}
\begin{figure}
    \centering
    \includegraphics[width=0.499\textwidth]{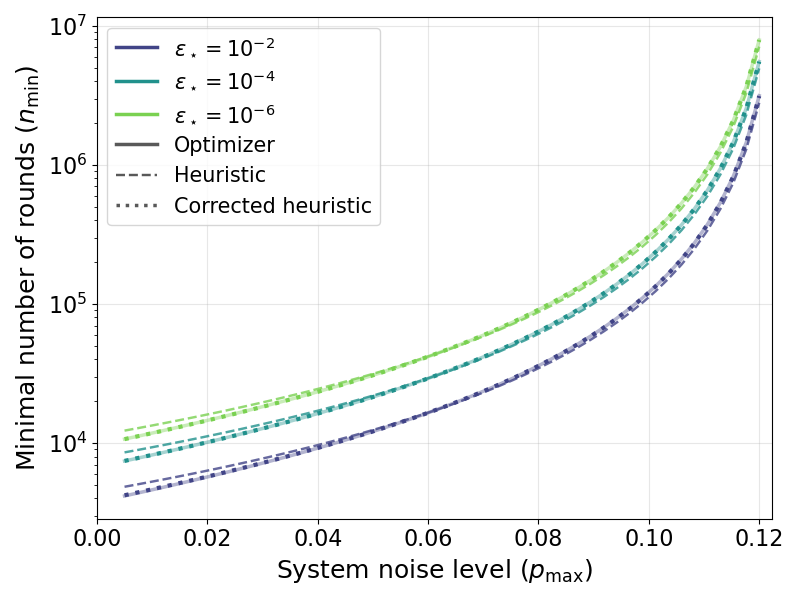}
    \caption{The scaling of $n_\text{min}$ with noise level $p_\text{max}$ for three different security targets $\epsilon_\star$ as given by the optimization (solid lines, made partially transparent to show overlaying dotted line) compared to the fitted heuristic function of Equation \ref{eq:heuristic_leading} (dashed lines) and corrected heuristic of Equation~\ref{eq:heuristic_corrected} (dotted lines). The heuristic model shows agreement with relative error of about 16\%, while the corrected heuristic shows agreement with relative error of about 1\%.}\label{fig:pmax-n}
\end{figure}

Having formalized the optimization problem, we now present its general solutions. By running the differential evolution optimizer across a range of operational scenarios, we determine the minimal number of rounds ($n_\text{min}$) required to reach a set security target. We also report on the operational parameters $\omega$ and $\tau$ that correspond to this $n_\text{min}$. The results quantify the resource cost imposed by both the physical hardware error rate (determining $p_\text{max}$) and the demand for strong security ($\epsilon_\star$).\\
The data in this section is generated for an inherent BQP error probability of $p=1/3$, corresponding to the standard definition of the BQP complexity class \cite{bernstein1993quantum}, and thus $Z=1/4$. We consider a $k=2$-colourable graph, which is the minimum possible value for $k$ and therefore maximizes the ratio $Z/k=1/8$, leaving as much room as possible for noise (because we need $p_\text{max}<Z/k$).\\
With $k$ and $p$ set, the only other input parameters are $\epsilon_\star$ and $p_\text{max}$. We will explore the effect of these parameters on the client's optimal strategy, 
which consists of $n_\text{min}, \omega$ and $\tau$.\\
First, we explore the effect of $p_\text{max}$ on $n_\text{min}$. In Figure \ref{fig:pmax-n} we show this behavior for $0 \leq p_\text{max} < Z/k$. Along with the optimization outcome, we plot two heuristic functions, which will get introduced in Section \ref{sec:heuristic}. We see that $n_\text{min}$ increases rapidly as $p_\text{max}$ approaches $Z/k$: the noise in the system almost saturates the probability of producing a wrong outcome even in an honest setting. This underlines the importance of keeping the noise in the system as low as possible, rendering the protocol practically not implementable for higher $p_\text{max}$.\\
The effect of $\epsilon_\star$ on $n_\text{min}$ is more subtle: while putting stronger requirements on the security does result in having to execute more rounds, as expected, this comes at a cost of more rounds that scales only linearly with $\log(\epsilon_\star)$.
We discuss these behaviors more when we introduce the heuristic function in Section \ref{sec:heuristic}.\\

\begin{figure*}[htbp]
    \centering
        \includegraphics[width=0.95\textwidth]{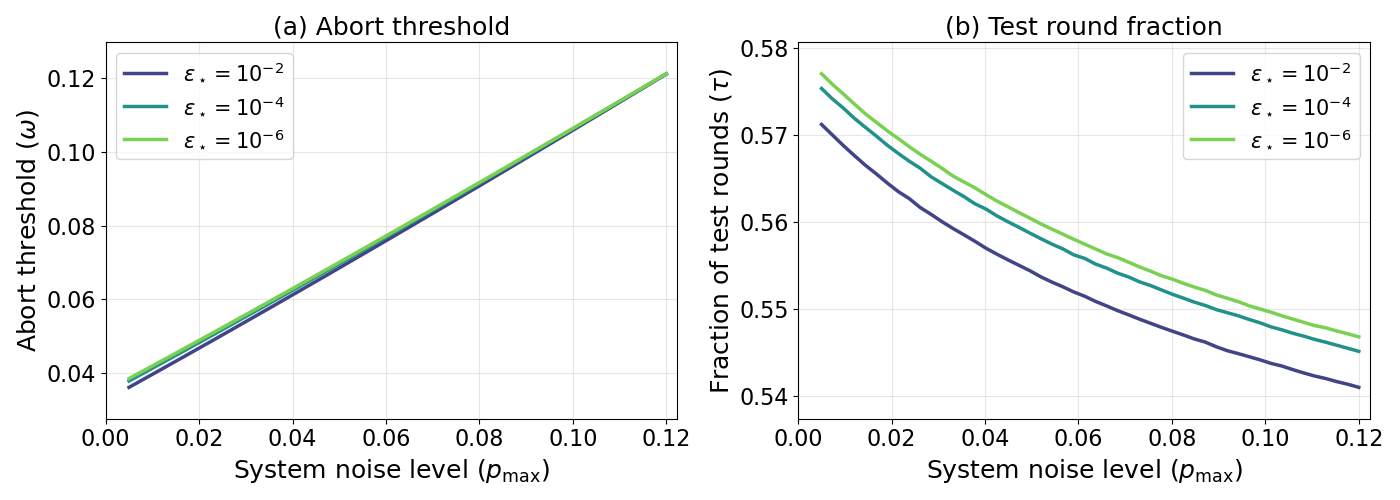}
        \caption{The scaling of protocol parameters (a) $\omega$ and (b) $\tau$ with noise level $p_\text{max}$ for three different security targets $\epsilon_\star$.}
    \label{fig:omega_tau}
\end{figure*}

Besides knowing how many total rounds to run the protocol for, it is also crucial to know how many of your $n_\text{min}$ rounds should be test rounds versus computation rounds ($\tau$), and how many test rounds are allowed to fail before aborting ($\omega$). Our optimization algorithm returns not only $n_\text{min}$ but also the corresponding optimized parameters $\phi, \epsilon_1, \epsilon_2, \epsilon_3$ and $\tau$, from which also $\omega$ can be determined using Equation \ref{eq:omega}.\\
The optimizer's role in finding both parameters, $\omega$ and $\tau$, is to find a balance point between opposing demands. In both cases, the values are largely determined by $p_\text{max}$, which tips the balance one way or the other, while $\epsilon$ has only a secondary effect. \\
For $\omega$, the allowed interval is ($p_\text{max}$, $Z/k$). If $\omega$ is set too low (too close to $p_\text{max}$), the protocol becomes oversensitive and will frequently abort due to normal hardware noise, compromising correctness. Conversely, if $\omega$ is set too high (too close to the theoretical limit $Z/k$), it gives a malicious server too much leeway to introduce errors without triggering the abort condition, compromising security. The optimizer places $\omega$ in between these extremes, at a point that can be written as $\omega = p_\text{max}+\alpha(Z/k-p_\text{max})$, for some $\alpha$. Empirically, $\alpha$ lies between about 0.20 and 0.26, showing a mild dependence on $\log(\epsilon_\star)$ and the ratio $p_\text{max}/(Z/k)$. The scaling of $\omega$ with $p_\text{max}$ is shown in Figure \ref{fig:omega_tau}(a).\\
For $\tau$ between 0 and 1, the optimizer balances between the number of test rounds and the number of computation rounds. More test rounds reduce the verification error $\epsilon_\text{ver}$, while more computation rounds reduce the correctness error $\epsilon_\text{cor}$. The trade-off is therefore how to allocate resources between verifying the server and reliably extracting the final outcome. We find again that this balance is mostly determined by $p_\text{max}$. Lower $p_\text{max}$ leads to a higher optimal $\tau$, since less noise means you can get a reliable answer to your computation with fewer computation rounds, leaving more rounds available for verification, this behavior can be seen in Figure \ref{fig:omega_tau}(b). The influence of $\epsilon_\star$ is weaker, since tightening the demand on $\epsilon$ increases demands on both verification and computation in similar proportion.

\subsection{Structure of the optimal strategy}
We find that the results that the optimizer returns have a common structure: At the optimum, the security exponents are equal ($A=B=C=D$), and the correctness and security requirements bind $n$ equally ($n_\text{cor}=n_\text{sec}$). Correctness and security compete for the same statistical budget, any margin spent on noise tolerance ($\omega-p_\text{max}$), is margin taken from security ($Z/k-\omega$) and vice versa. We find that both equalities hold true to numerical precision across the range of $p, k, \epsilon_\star$ and $p_\text{max}$ tested. While this simplifies the optimization, we do not implement this structure into the solver, as this is only an empirical structure. It does, however, motivate the heuristic function we discuss below. 

\subsection{Heuristic for minimal number of rounds}\label{sec:heuristic}
The structure of the optimum gives us the scaling of $n_\text{min}$ directly. The noise-tolerance constraint $p_\text{max}<\omega<Z/k$ means the statistical budget is the gap $Z/k-p_\text{max}$. As $p_\text{max}\rightarrow Z/k$ this statistical budget becomes smaller, and there is less slack. The tuning parameters $\phi, \epsilon_1, \epsilon_2, \epsilon_3$ are each bounded by the gap and vanish linearly with it. Correctness is set by the Hoeffding exponent $(\omega - p_\text{max})^2$, the security is set by $\min(A,B,C,D)$. Each of the exponents is quadratic in a tuning parameter, meaning both $n_\text{cor}$ and $n_\text{sec}$, and thus $n_\text{min}$, scale like $(\omega-p_\text{max})^{-2}$. 

In addition, we find a log-dependence on the security target $\epsilon_\star$. The logarithmic dependence comes from two parts: Solving the equality \ref{eq:ncor} gives a term proportional to $-\log(\epsilon_\star)$, and the security equality (Equation~\ref{eq:nsec}) contributes $\log(64/\epsilon_\star^2)=-2\log(\epsilon_\star)+\log(64)$. We find that at the optimum, these are equal, so the linear dependence of $n_\text{min}$ on $\log(\epsilon_\star)$ is justified. 

We combine these findings into a leading-order heuristic
\begin{equation}
    n_{\min} \approx \frac{-a\log\epsilon_\star + b}{\left(Z/k - p_{\max}\right)^{2}}.
    \label{eq:heuristic_leading}
  \end{equation}
Fitting this to the output of the optimizer gives $a=11.6$, $b=16.3$. Figure \ref{fig:pmax-n} shows good agreement between this heuristic and the optimizer data. Taking a closer look at the accuracy of the heuristic we find that the relative error goes up to at most about 16\% in the extreme values of $p_\text{max}$ ($p_\text{max}\rightarrow 0$ or $p_\text{max}\rightarrow Z/k$). A 16\% error might seem large at first glance, but this parameter space spans many orders of magnitude, so being within 16\% gives a good estimate of the required resources. With this, our heuristic model can serve as a powerful tool for rapid resource estimation without needing the full optimization as well as provide a better understanding of the behaviour. This heuristic function shows the two fundamental costs of the protocol:
\begin{enumerate}
    \item The cost of security: the resource scales linearly with -log($\epsilon_\star$). This means that each additional order of magnitude in security (adding a `nine' of security) comes at a fixed, additive cost in the number of rounds. 
    \item The cost of noise: the resource cost follows a power-law divergence as the hardware noise approaches it tolerance limit. This implies $p_\text{max}$ should be significantly lower than $Z/k$ for the protocol to be practically implementable.
\end{enumerate}

We find that, while this heuristic reproduces the dominant behavior, a slow drift of the values $a$ and $b$ remains as the noise level changes. We find that this drift is almost entirely captured by the ratio $r\equiv p_{\max}/(Z/k)$. We absorb this drift into a low-order correction, of the form
\begin{equation}
    n_{\min} \approx \frac{-a\log\epsilon_\star + b}{\left(Z/k - p_{\max}\right)^{2}}
    \left(1 + c_1\, r + c_2\, r^2\right).
    \label{eq:heuristic_corrected}
  \end{equation}
Using $Z/k - p_{\max} = (Z/k)(1-r)$. This correction becomes relevant when $r\to 0$, where the tuning parameters are limited by their absolute bounds. Fitting this form gives $a\approx9.95, b\approx13.99, c_1\approx0.37$ and $c_2\approx0.06$, this reduces the relative error from $~16\%$ to $~1\%$. While this form is less elegant, it provides a very good estimate of the required resources.

\part{Case study}
To demonstrate the practical utility of our optimization framework, we apply it to a realistic experimental scenario: a measurement-only client connected to a trapped-ion quantum server executing a minimal computation using the rVBQC protocol. This case study serves two purposes. First, it provides concrete resource estimates for a near-term experimental implementation of the rVBQC protocol. Second, it illustrates how the framework can be used to optimize system-level design choices, specifically the rate-fidelity trade-offs inherent in quantum network architectures.\\
We will first introduce the general setup and the trade-off we want to study. We then provide the optimal protocol parameters for this setup. Finally, we present results that can guide experimentalists in hardware trade-offs, demonstrating how the framework can be used more broadly.

\section{Setup}\label{sec:setup}
\subsection{System architecture}\label{subsec:sysarch}
We consider a configuration where a resource-limited client, capable only of rotating, and measuring single photonic qubits, delegates computation to a remote trapped-ion quantum processor, a setup as described in e.g. Ref. \cite{drmota2024verifiable}. Previous work also simulated this setup~\cite{van2024hardware}, and minimized the hardware cost to make the rVBQC protocol possible at all, by setting $\omega$ to its maximum of 25\%. Here, instead, we assume the hardware is good enough to run the protocol, and we aim to balance certain harware parameters in order to yield the shortest end-to-end runtime. To satisfy the hardware constraints imposed by current technology while remaining within the protocol's noise tolerance ($p_\text{max} < Z/k$), we implement the simplest possible blind computation: a two-qubit linear cluster state, which has $k = 2$ colorability. We assume a distance between the client and server of 10km. \\
For the case study, we consider a deterministic computation ($p=0$), yielding a more relaxed noise threshold of $Z/k = 1/4$. This allows us to demonstrate the framework using hardware parameters representative of current technology. For BQP computations with $p=1/3$, the threshold tightens to $Z/k = 1/8$, which remains beyond current capabilities.\\
The rate-fidelity trade-off we study here is a hardware decision: the client device should rotate and measure incoming photonic qubits. Since we require fresh randomness for each attempt, the client should be able to rotate the incoming photon, in polarization basis, for each attempt. If the re-configuration of the polarization controllers takes longer than the time it takes to emit a photon from the ion plus the time it takes the photon to travel to the client, the time per attempt ($t_\text{round}$) will take longer (higher $t_\text{round}$). However, in current hardware having faster control often comes at a cost of having lower control precision, leading to more errors (higher $p_\text{max}$) and therefore more rounds to be performed (higher $n_\text{min}$). The end-to-end runtime ($=n_\text{min}(p_\text{max})t_\text{round}$) is the only relevant metric, as security, rate and fidelity are covered by this. Any decision on the hardware should therefore be based on what minimizes the runtime for a given $\epsilon_\star$. 

\subsection{Hardware model and noise sources}
To apply our framework to the polarization controller trade-off, we need to understand the effect of the trade-off on $p_\text{max}$ and $t_\text{round}$. We characterize the polarization controller by two parameters: the polarization control error, and the polarization control delay. The control error leads to a random deviation in the angles that define the polarization state. Polarization controllers are typically characterized by their mechanical repeatability, which is specified by the manufacturer to indicate how reproducibly the device can reach a target setting. Due to the geometry of polarization optics, a physical rotation of the controller by angle $\theta$ results in a rotation of the polarization state on the Bloch sphere by $2\theta$. We therefore model the control error by sampling angular deviations from a normal distribution with standard deviation equal to twice the mechanical repeatability, and adding these deviations to the polar and azimuthal angles of the client's intended measurement basis.\\
 The polarization delay affects the computation only when it is longer than the communication time between the client and server plus the time it takes the server to initialize its qubit and emit a photon to the fiber. When it takes longer than that, it will increase the time per remote state preparation attempt. We require multiple remotely prepared states per round, so the second state needs to be remotely prepared before a cutoff time that depends on the coherence time of the server qubits. Because of this, the effect on $t_\text{round}$ is more than linear.\\
To find the effect of the different polarization controller hardware on the protocol we simulate the execution of a test round using the quantum network simulator NetSquid \cite{coopmans2021netsquid}. We find $p_\text{max}$ by averaging the fraction of tests that failed and $t_\text{round}$ by averaging the time it takes to complete the test round.\\
Besides the effects of the polarization control hardware, we include the following noise models:
\begin{enumerate}
    \item Collective dephasing of the trapped ions with a coherence time of 4 seconds \cite{innsbruck_params}
    \item Probability of depolarizing trapped ion when applying a single-qubit gate of 0.01 \cite{krutyanskiy2023telecom}
    \item Probability of depolarizing trapped ion when applying a two-qubit gate of 0.05 \cite{krutyanskiy2023telecom}
    \item Ion-photon entanglement as a Werner state with fidelity 0.88 \cite{krutyanskiy2023entanglement}
    \item Dark count probability at the client detecors as depolarizing noise with depolarizing probability 0.02 \cite{innsbruck_params}
    \item Measurement error probability of ion trap of 0.0001 \cite{innsbruck_params}
    \item Loss modes: server efficiency (including coupling to fiber \cite{schupp2021interface} and frequency conversion \cite{innsbruck_params}) of 0.287, probability of successfully transmitting through fiber of $10^{-10*0.2/10}\approx0.63$, detector efficiency of 0.87 \cite{schupp2021interface}.
\end{enumerate}
In the code, we made several simplifications that would lead to an incorrect implementation of the protocol, but lead to a correct estimation of $p_\text{max}$ and $t_\text{round}$. For example, in the simulation the client does not use a one-time pad or randomly chosen measurement angles for its remote state preparation. Instead, it always applies the same rotation. This simplifies the classical bookkeeping but does not affect the quantum errors that might occur in the protocol (at least not those covered by the simulation).\\

\section{Results}\label{sec:results}
\begin{figure}
    \centering
    \includegraphics[width=0.499\textwidth]{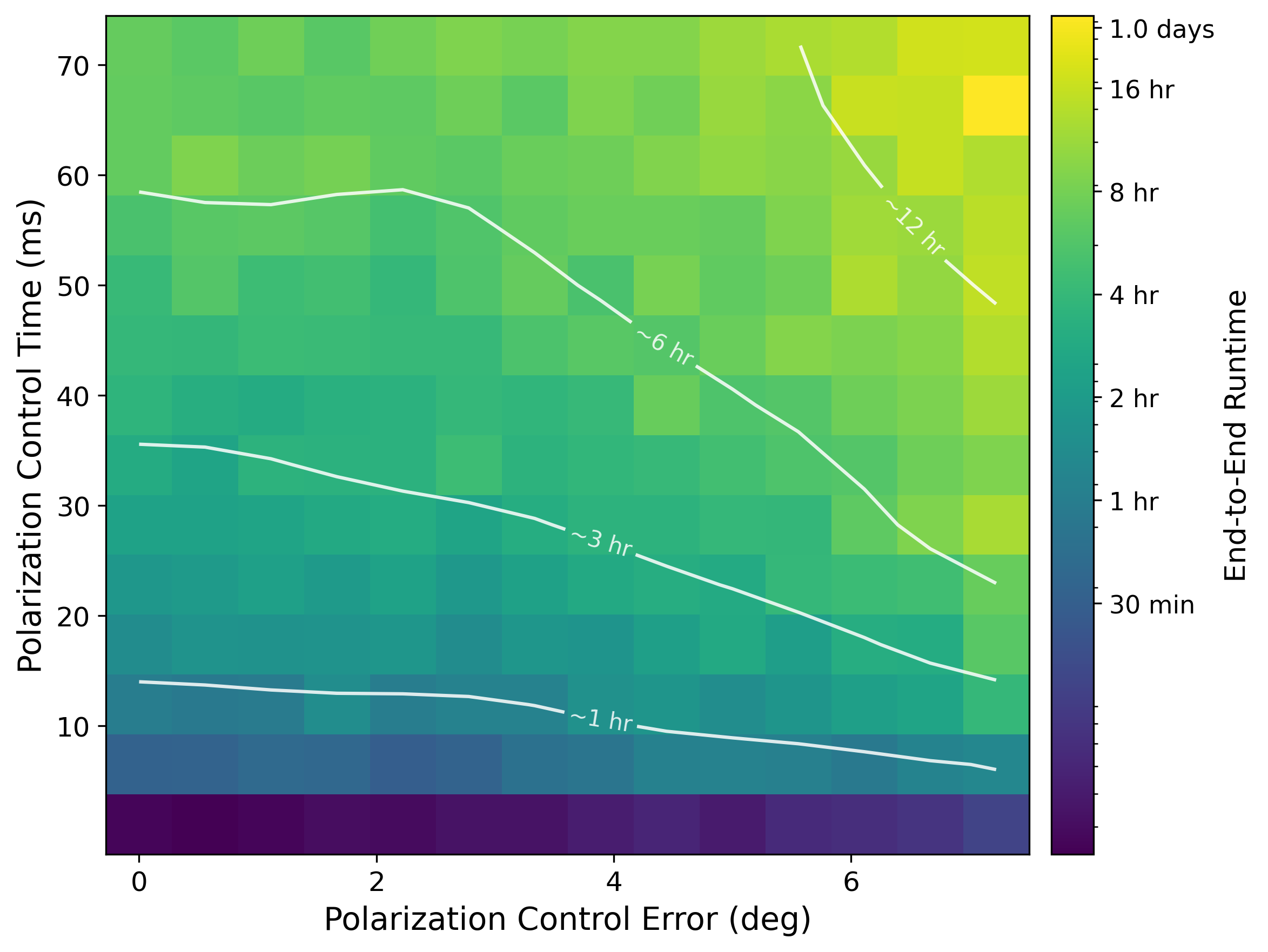}
    \caption{End-to-end protocol runtime as a function of polarization controller parameters. Color indicates runtime on a logarithmic scale. White contours mark approximate iso-runtime boundaries (Gaussian-smoothed, $\sigma=1.2$ grid cells).}\label{fig:end-to-end}
\end{figure}

We simulate the protocol for different combinations of polarization control parameters (time and precision), running 10000 iterations at each point to estimate $p_\text{max}$ and $t_\text{round}$. With this, we run the optimization framework to find the corresponding $n_\text{min}$.\\
The resulting end-to-end runtime ($n_\text{min} \cdot t_\text{round}$) is shown in Figure \ref{fig:end-to-end} as a function of polarization control error and polarization control time for security target $\epsilon_\star=10^{-4}$. The color scale is logarithmic, spanning approximately 6 minutes to 1 day. White contour lines indicate approximate iso-runtime boundaries at 1, 3, 6, and 12 hours, smoothed to reduce noise from finite sampling.\\
The contours in the low-error, low-time regime (bottom-left) are predominantly horizontal, suggesting that in this regime the end-to-end runtime is primarily limited by control time rather than control error. This is consistent with the behavior predicted by our optimization framework: when hardware noise is well below the protocol's tolerance threshold ($p_\text{max}\ll Z/k$), the number of rounds $n_\text{min}$ remains relatively constant, and the total runtime scales with $t_\text{round}$.\\
The contours in the high-error, high-time regime (top right) start to curve upward, indicating that control error starts to have a stronger effect on the runtime.\\
These results carry a practical implication for hardware selection: in the regime we studied, the client should predominantly prioritize fast switching over high angular precision.

\section{Conclusion}
We have presented a practical framework for implementing the noise-robust verifiable blind quantum computing protocol of Leichtle et al. The protocol's security proof involves numerous interdependent parameters, making it non-trivial to find a valid parameter set for given hardware noise and security requirements. We formalized a constrained optimization problem to address this issue and developed a method to solve it using differential evolution, yielding the protocol parameters that minimize the number of rounds for any given setup. As a secondary result, we derived a heuristic formula for rapid resource estimation, showing that the round cost scales logarithmically with the security parameter and diverges as a power law when hardware noise approaches the protocol's tolerance threshold. We also provide a corrected heuristic that, while less intuitive, provides an even better estimate with a residual error of only about 1\%.\\
Since the number of rounds depends on noise while the time per round depends on hardware rate, our framework enables optimization of rate-fidelity trade-offs to minimize end-to-end runtime. We demonstrated this through a case study of a trapped-ion server with a measurement-only client, analyzing the trade-off between polarization control precision and switching speed. The results show that, in the regime where hardware noise is well below the protocol's tolerance, fast switching should be prioritized over high precision—an encouraging finding for near-term implementations.\\As hardware improves and the more stringent noise thresholds required for general BQP computations become achievable, the methodology presented here will provide concrete guidance for experimental implementations of secure delegated quantum computation.

\section{Data and code availability}
The code used to generate, process and plot the data is available at \cite{code}. The main dataset for the case study (Figure \ref{fig:end-to-end}) is published along with the code. 

\section*{Acknowledgements}
The authors would like to thank Sounak Kar for pointing to useful bounds, and Guus Avis, Jeroen Grimbergen, Sergio Loarte and Harold Ollivier for useful feedback on the manuscript and discussions. The authors would additionally like to thank Ben Lanyon and Tracy Northup and groups for providing information about their trapped-ion systems.\\
This project has received funding from the European Union’s Horizon Europe research and innovation programme under grant agreement No. 101102140.

\section*{Author contributions}
JvD executed the project and wrote the manuscript. MvH provided assistance in simulation efforts and code maintenance. SW supervised the project. 

\bibliographystyle{unsrt}
\bibliography{bib}

\clearpage
\appendix
\section{Glossary}\label{app:glossary}
Here we provide an overview of the parameters introduced in this paper, for reference. We divide the parameters up in the following categories:
\begin{enumerate}
    \item Input parameters: these parameters define the optimization problem and are user-provided, these are presented in Table \ref{tab:input_paramsapp}
    \item Tuning parameters: parameters that are part of the security proof and can be tuned the find the minimal number of rounds. These are presented in Table \ref{tab:tuning_paramsapp}.
    \item Other parameters: parameters that occur but do not fit into the above two categories. These are presented in Table \ref{tab:other_paramsapp}
\end{enumerate}

\begin{table}[!htb]
    \centering
    \caption{Input parameters}
    \label{tab:input_paramsapp}
    \begin{tabular}{|p{0.8cm}|p{5.5cm}|}
        \hline
        & \textbf{Meaning} \\
        \hline
        $\epsilon_\star$ & Overall security error target\\
        \hline
        $p_\text{max}$ & The probability that a test round fails in an honest-but-noisy setting: determined by the physical noise level of the system\\
        \hline
        $k$ & The colourability of the graph that is used in the computation. This tells us the minimal number of colours needed to colour a graph such that no adjacent vertices share the same colour\\
        \hline
        $p$ & The maximum probability that a quantum algorithm in the bounded-error quantum polynomial time class can give an incorrect answer on any given input\\
        \hline
    \end{tabular}
\end{table}

\begin{table}[!htb]
    \centering
    \caption{Tuning parameters}
    \label{tab:tuning_paramsapp}
    \begin{tabular}{|p{0.8cm}|p{5.5cm}|}
        \hline
        & \textbf{Meaning} \\
        \hline
        $\phi$ & Defines the threshold for the fraction of total rounds affected, which determines when a cheating strategy is ``conservative" or ``aggressive"\\
        \hline
        $\epsilon_1$ & Used to bound the probability of the number of affected test rounds being significantly lower than expected\\
        \hline
        $\epsilon_2$ & Buffer used in setting the round failure threshold\\
        \hline
        $\epsilon_3$ & Bounds fluctuations in how server deviations distribute across computation rounds \\
        \hline
        $\tau$ & Fraction of test rounds over the total number of rounds\\
        \hline
    \end{tabular}
\end{table}

\begin{table}[!htb]
    \centering
    \caption{Other parameters}
    \label{tab:other_paramsapp}
    \begin{tabular}{|p{0.9cm}|p{5.4cm}|}
        \hline
        & \textbf{Meaning} \\
        \hline
        $G=(V,E)$ & Description of a graph with vertices $V$ and edges $E$, used to define a graph state for the MBQC computation\\
        \hline
        $d$ & Number of computation rounds in the protocol\\
        \hline
        $t$ & Number of test rounds in the protocol\\
        \hline
        $\epsilon_\text{cor}$ & Local correctness error\\
        \hline
        $\epsilon_\text{bl}$ & Local blindness error\\
        \hline
        $\epsilon_\text{ver}$ & Local verifiability error\\
        \hline
        $\epsilon_\text{ind}$ & Independent verification error\\
        \hline
        $\epsilon_\text{rej}$ & Probability of aborting due to honest noise\\
        \hline
        $\epsilon_4$ & Relates inherent BQP error $p$ to the threshold for corrupted computation rounds\\
        \hline
        $\omega$ & Fraction of test rounds the client allows to fail before aborting the protocol\\
        \hline
        $Z$ & Shorthand notation for $Z=(2p-1) / (2p-2)$\\
        \hline
        Pr[fail] & Probability that a cheating server gets away with providing a wrong answer, set by Equation \ref{eq:prfail}\\
        \hline
    \end{tabular}
\end{table}

\end{document}